\documentclass[aps,tightenlines,nofootinbib,preprint,groupedaddress]{revtex4}
\usepackage{graphicx}
\newcommand{\be}{\begin{equation}}
\newcommand{\ee}{\end{equation}}
\newcommand{\bear}{\begin{eqnarray}}
\newcommand{\eear}{\end{eqnarray}}
\begin{document}

\title{Nonperturbative effects from the resummation of \\ perturbation theory}

\author{Taekoon Lee}
\email{tlee@muon.kaist.ac.kr}

\affiliation{Department of Physics, Korea Advanced Institute of
Science and Technology, Daejon 305-701, Korea}


\begin{abstract}

Using the general argument in Borel resummation of
perturbation theory that links the divergent perturbation theory to the
nonperturbative effect we argue that the nonperturbative effect
associated with the perturbation theory should have
a branch cut only along the positive real axis in the complex coupling plane.
The component in the weak coupling expansion of the
nonperturbative amplitude  that gives rise to the
branch cut can be calculated in principle from the perturbation theory 
combined with the exactly calculable properties of the nonperturbative effect.
The realization of this mechanism is demonstrated in the double well potential
and the two-dimensional $O(N)$ nonlinear sigma model. In these models the
leading term in the weak coupling expansion
of the nonperturbative effect can be obtained
with a good accuracy from the first terms of  perturbation theory.
Applying this mechanism to the infrared renormalon
induced nonperturbative effect in QCD, we suggest some of 
the QCD condensate effects can be calculated in principle from the
perturbation theory.
\end{abstract}



\maketitle

\section{\label{}Introduction}
The usual perturbation  in the weak coupling constant in field
theory is an asymptotic expansion. When the perturbation series 
is sign alternating 
it may be resummed, for example, in the manner of
Borel resummation. However, when the
series is not sign alternating, it usually implies the presence of a genuine
nonperturbative effect, and the Borel resummation of the perturbation 
series alone is in principle not sufficient for an adequate description of the
true amplitude.

The nonperturbative effect and the perturbation expansion are not totally
independent; the former controls the large order behavior of the latter.
Even with this relation, however, explicit calculations show that the 
nonperturbative effect cannot be calculated from the perturbation series
even when the latter is known to all orders.

The purpose of this paper is to argue, by taking a closer look at
the general but heuristic argument that relates
the nonperturbative effect with the perturbation theory, that some parts
of the nonperturbative effect, which usually include 
the leading piece in weak coupling expansion,
can be calculated in principle from the Borel resummation of perturbation
theory along with 
the exactly calculable properties of the nonperturbative effect.

\section{\label{sec2} Calculable component in nonperturbative effect}

A general but heuristic argument that relates the nonperturbative
effect with perturbation theory goes as follows.\footnote{
A good introduction can be found in \cite{justin0,justin1}.}
Let $A(\alpha)$
be an amplitude with perturbation expansion in the coupling constant
$\alpha$:
\be
A(\alpha)=\sum_0^\infty \, a_n \,\alpha^{n+1}
\label{series}
\ee
and assume that $A(\alpha)$ be real for $\alpha > 0$.
We shall further assume that $a_n$ at large orders is nonalternating
in sign. In general, $a_n$ diverges factorially due to renormalons or
instanton--anti-instanton pairs, and in principle 
Eq. (\ref{series}) is meaningless
unless some kind of resummation is performed on the divergent series.
To do a resummation, consider a new series
\be
B(\alpha)=\sum_0^\infty \, a_n\, (-1)^{n+1}\,\alpha^{n+1}
\label{series2}
\ee
which is obtained from Eq. (\ref{series}) by flipping the sign of
the coupling constant.
Since for $\alpha >0$ this series is alternating in sign, it can now be 
Borel resummed, yielding a resummed amplitude $B_{\rm PT}(\alpha)$,
\be
B_{\rm PT}(\alpha)\equiv\int_0^\infty db \,
e^{-\frac{b}{\alpha}} \tilde B(b) 
\label{borel}
\ee
where
\be
\tilde B(b) =\sum_0^\infty \frac{(-1)^{n+1} \, a_n}{n!}\, b^n \, .
\ee

One would expect that $A(\alpha)$ could be obtained from the
Borel resummed $B_{\rm PT}(\alpha)$ by analytic continuation from the
positive real axis to the negative real axis in the complex $\alpha$-plane.
The problem is, however, $B_{\rm PT}(\alpha)$ is expected
to have a branch cut along the negative axis, and consequently
$B_{\rm PT}
(- \alpha \pm {\rm i} \epsilon) $ will have an imaginary part.
Therefore analytic continuation alone of $B_{\rm PT}
(\alpha)$ cannot reproduce
$A(\alpha)$ which is by definition real for $\alpha > 0$.
The expected resolution of this problem is that the true amplitude
$A(\alpha)$ has a nonperturbative
amplitude  $A_{\rm NP}(\alpha)$ in addition to the
$B_{\rm PT}(-\alpha)$, so when they are added together the  imaginary
parts from each amplitude cancel each other, rendering the total
amplitude to be real.
That is,
\be
A(\alpha)=A_{\rm PT}(\alpha) + A_{\rm NP}(\alpha)
\label{master}
\ee
with
\be
{\rm Im}\, [A_{\rm PT} (\alpha \pm {\rm i} \epsilon) ]+
{\rm Im}\,[ A_{\rm NP} (\alpha \pm {\rm i} \epsilon )] =0 \, ,
\label{imaginary}
\ee
where $A_{\rm PT}(\alpha)\equiv B_{\rm PT}(-\alpha)$,
which may be 
called the perturbative amplitude, has now a branch cut  along the positive
real axis in  the $\alpha$-plane.
Performing the analytic continuation in Eq. (\ref{borel})
we obtain
\be
A_{\rm PT} (\alpha \pm {\rm i} \epsilon) =
\int_{0\pm {\rm i} \epsilon}^{\infty \pm {\rm i} \epsilon}db\,
e^{-\frac{b}{\alpha}}
\tilde A (b)\,
\label{alpha-b-relation}
\ee
where\footnote{Throughout the article 
whenever a Borel transform is defined through a perturbation series
like (\ref{borel-series2}) it is assumed that the value of the Borel transform
at a point beyond the convergence disk is obtained by analytic continuation.}
\bear
\tilde A (b)&=&- \tilde B(-b) \nonumber \\
&=& \sum_0^\infty \frac{a_n}{n!} b^n \,.
\label{borel-series2}
\eear
Note that, as we make  analytic continuation of $B_{\rm PT}(\alpha)$
from the positive real axis to the negative real axis in the $\alpha$-plane
counterclockwise (clockwise)
in the upper (lower) half plane, the integration contour also should rotate
counterclockwise (clockwise), hence the 
$\pm$ sign in Eq. (\ref{alpha-b-relation}).

Using the dispersion relation on the resummed $A_{\rm PT}(\alpha)$ and
Eq. (\ref{imaginary}), we have
\bear
A_{\rm PT}(\alpha)&=&\frac{1}{\pi}\int_0^\infty
\frac{{\rm Im}[ A_{\rm PT}(\alpha' +{\rm i}\epsilon)]}{
\alpha'-\alpha}\,d\alpha' \nonumber\\
&=&-\frac{1}{\pi}\sum_{n=0}^\infty \left[\int_0^\infty \frac{{\rm Im}
[A_{\rm NT}(\alpha' + {\rm i}\epsilon)]}{\alpha'^{n+2}}\right]
\,\alpha^{n+1}\,.
\eear
Thus the perturbative coefficient in Eq. (\ref{}) can be written as 
\be
a_n= -\frac{1}{\pi} \int_0^\infty \frac{
{\rm Im}\, 
[A_{\rm NP} (\alpha + {\rm i} \epsilon )]}{\alpha^{n+2}}\,d \alpha
\label{dispersion} \,
\ee
which makes the relation between the nonperturbative effect
and the perturbation theory explicit.
At large values of $n$ the dominant contribution
in (\ref{dispersion}) comes from the small $\alpha$ region, and so
for large order behavior only the weak coupling limit of the
nonperturbative effect is required.

Now with Eqs. (\ref{master}) and (\ref{imaginary})  
the  amplitude $A(\alpha)$ for $\alpha >0$ 
can be written as the sum of the real parts of the
perturbative and the nonperturbative terms:
\be
A(\alpha)={\rm Re}\, [A_{\rm PT} (\alpha \pm {\rm i} \epsilon) ]+
{\rm Re}\, [ A_{\rm NP} (\alpha \pm {\rm i} \epsilon )]\,.
\ee
Equation (\ref{imaginary}) also shows that the imaginary part of 
the nonperturbative amplitude $A_{\rm NP}(\alpha)$
can be calculated in principle from the Borel resummation of the
perturbation theory. 
The real part, however, is in general not calculable from 
the perturbation theory.

The argument hitherto is well known, perhaps except for
Eq.~(\ref{alpha-b-relation})  which allows us to relate  the imaginary
part from the analytic continuation in the $\alpha-$plane to 
that arising from the Borel integral.
Now, our observation, which will play a crucial role 
throughout the paper, is that Eq. (\ref{imaginary}) suggests
$ A_{\rm NP} (\alpha)$ have a branch cut along the
positive real axis in the complex coupling plane, in order to cancel the
imaginary part coming from the branch cut in the Borel resummed
perturbative amplitude.
This rather straightforward observation can have an important consequence:
it renders some part of the nonperturbative amplitude to be
calculable from the perturbation theory.\footnote{The idea of rebuilding
nonperturbative effects from perturbation theory is an old story.
To the author's knowledge it
was first speculated in \cite{david1}, and an approach similar to ours was
observed in \cite{grunberg}.}
In the weak coupling limit ($\alpha \to 0$) the approximate
functional form of $A_{\rm NP}(\alpha)$ can be rather easily
determined from
other nonperturbative techniques such as renormalization group
argument or instanton calculations, and the component, typically
the leading one
in weak coupling expansion, that could
give rise to a branch cut along the positive real axis can
be easily identified. We then further fix this component to a more
specific form by
{\it demanding that it give a branch cut only along the positive real axis;}
We expect the nonperturbative effect should not have other branch cut, 
for instance,
such as one along the negative real axis, since it would imply
that the perturbation series (\ref{series}) is not Borel-summable
even for $\alpha <0$.
The constraint on the functional form from this step turns out 
to be sufficient enough for us to relate  the real
part of the above-mentioned component to its imaginary part,
consequently rendering the former to be calculable from the Borel
resummation of the perturbation theory through 
Eq. (\ref{imaginary}). 

In the next two sections we consider a few definite examples
and show how this procedure can be realized in model calculations.
In these examples we shall focus on the nonperturbative effect
associated with the  first singularity 
on the positive real axis in the Borel plane.

\section{The double well potential}

The quantum mechanical double well potential with the action
\be
S=\int \left[ \frac{1}{2} \dot{q}^2 -\frac{1}{2} q^2 (1-\lambda q)^2\right]
dt
\ee
has instanton solutions, and the nonperturbative effects due to the instantons
give rise to singularities on the positive real axis on the Borel plane,
causing same sign perturbation series.

Consider, for example, $E(\alpha)$ defined by
\be
E(\alpha)= \frac{1}{2}\left[ E_+(\alpha)+E_-(\alpha)-1\right]\,
\ee
where $\alpha\equiv \lambda^2$ and $E_-(\alpha),\, E_+(\alpha)$
are, respectively, the energies of the ground and the first excited states.
The Borel transform  
\be
\tilde E(b) =\sum_{n=0}^\infty \frac{a_n}{n!} b^n
\ee
of the perturbation series for $E(\alpha)$,
\be
E(\alpha) =\sum_{n=0}^\infty a_n \alpha^{n+1},
\label{e-series}
\ee
is expected to have multi-instanton caused singularities at
$b=2 n S_0$, where $ n=1,2,3,\cdots$, and 
$S_0=1/6$ is the one-instanton action.

The large order behavior of the perturbation (\ref{e-series})
is controlled by the first singularity at $b=1/3$.
The nonperturbative effect that causes this singularity is due 
to the contributions of instanton--anti-instanton pairs, and can be calculated 
from the potential of an instanton--anti-instanton pair,
and reads \cite{justin2}
\be
E_{{\rm NP}}(\alpha) = \frac{1}{\pi\alpha}e^{-1/3\alpha}\left[
\left(\ln\left(\frac{-2}{\alpha}\right)+\gamma_{\rm{E}}
\right)\left(1-\frac{53}{6}\alpha\right)
-\frac{23}{2} \alpha +O(\alpha^2 \ln \alpha)\right]\,
\label{ins-eng}
\ee
where $\gamma_{\rm{E}}$ is the Euler constant.

Notice that $E_{{\rm NP}}(\alpha)$ has a
branch cut along the positive real axis in the $\alpha$-plane,
in agreement with our argument in the previous section.
The minus sign in the logarithmic term, which causes the branch cut,
arises from the required sign flip in the coupling constant
to pick up the nonperturbative effect from the attractive 
potential  of the instanton--anti-instanton pair \cite{bogom}.

The imaginary and the real parts of $E_{{\rm NP}}(\alpha)$ now read 
\bear
{\rm Im} [E_{{\rm NP}}(\alpha \pm {\rm i} \epsilon)]
&=&\pm \frac{1}{\alpha} e^{-1/3\alpha}\left[ 1+ \frac{53}{6} \alpha
+O(\alpha^2)\right] \label{imaginarypart} \\
{\rm Re} [E_{{\rm NP}}(\alpha\pm {\rm i} \epsilon)]&=&
\frac{1}{\pi\alpha}e^{-1/3\alpha}\left[
-\ln(\alpha)+ \ln(2)+\gamma_{\rm{E}}
 +O( \alpha)\right]\label{realpart}\,.
\eear
Note that the real part has terms, for example, the constant terms within
the bracket in Eq. (\ref{realpart}),
that have nothing to do with
the imaginary part. These terms represent genuine nonperturbative 
effect and cannot be calculated from Borel resummation of
perturbation theory.

Since in this example the nonperturbative effect can be calculated
in weak coupling expansion there is no real need to attempt to 
extract the nonperturbative effect from the perturbation theory.
However, for the sake of argument, let us suppose that
we knew only that the nonperturbative effect in weak coupling expansion
was given in the form:
\be
E_{{\rm NP}}(\alpha)\propto\frac{1}{\alpha} e^{-1/3\alpha}\left[
\ln (\alpha)+{\rm subleading~ terms}\right] \, .
\label{roughform}
\ee
In any event, inferring this form may not be so difficult since the
preexponential factor $1/\alpha$ can be obtained by counting the
number of
(quasi)zero modes, in this case two, and the logarithmic term arises
from the instanton--anti-instanton potential at large distance.

We can now improve the form of the nonperturbative effect (\ref{roughform})
by demanding that it have a branch cut only along the
positive real axis in the $\alpha-$plane.
Because the branch cut can arise only from the logarithmic term
we can immediately see that $E_{{\rm NP}}(\alpha)$ must assume 
the following form:
\be
E_{{\rm NP}}(\alpha)=-\frac{C}{\pi\alpha} e^{-1/3\alpha}\left[
\ln (-\alpha)+{\rm subleading~ terms}\right] \, 
\label{improvedform}
\ee
with $C$ an unknown real constant.
Of course, a comparison with Eq. (\ref{ins-eng}) shows that
the true value of the constant must be $C=1$.
We now show that the leading term, the logarithmic term, in the real part
(\ref{realpart}) can be calculated from the perturbation theory
starting from the ansatz (\ref{improvedform}).
The imaginary and the real parts from this expression (\ref{improvedform})
are then
\bear
{\rm Im} [E_{{\rm NP}}(\alpha \pm {\rm i} \epsilon)]
&=&\pm \frac{C}{\alpha} e^{-1/3\alpha}\left[ 1+ {\rm subleading~ terms}
\right] \label{imaginarypart-1} \\
{\rm Re} [E_{{\rm NP}}(\alpha\pm {\rm i} \epsilon)]&=&
-\frac{C}{\pi\alpha}e^{-1/3\alpha}\left[
\ln(\alpha)+ {\rm subleading~ terms}\right]\label{realpart-1}\,.
\eear
To determine the leading term in the real part we now only need
to fix the constant $C$.
This constant becomes the residue, up to a calculable normalization,
of the first singularity in the
Borel plane, and can be calculated in perturbation theory \cite{tlee1,tlee2}.
In fact,  for the Borel resummation
\be
E_{\rm PT} (\alpha \pm {\rm i} \epsilon) =
\int_{0\pm {\rm i} \epsilon}^{\infty \pm {\rm i} \epsilon}
e^{-\frac{b}{\alpha}}
\tilde E (b)\, db
\label{borelsum}
\ee
to have imaginary parts that can cancel the imaginary terms
in Eq. (\ref{imaginarypart-1})
the Borel transform $\tilde E(b)$  must have a singularity
at $b=1/3$ of the form
\be
\tilde E(b)= -\frac{9C}{\pi(1-3 b)^2} [1+ O(1-3b)]\,.
\ee

To determine the residue we now consider a function $R(b)$ defined by
\be
R(b)=\tilde E(b) (1-3b)^2 \,.
\ee
The difference between $R(b)$ and $\tilde E(b)$ is that the former has
a much softer singularity. Although it may appear that $R(b)$ is regular
at $b=1/3$, it is easy to see that that is not the case.
In fact, for the imaginary parts from the 
Borel integral (\ref{borelsum}) to cancel the
imaginary parts in Eq. (\ref{imaginarypart}),
$\tilde E(b)$ should have an expansion
around the singularity 
\be
\tilde E(b)=- \frac{9}{\pi(1-3b)^2}
\left[ 1 -\frac{53}{18}(1-3b) +O[ (1-3b)^2\ln(1-3b)]\right]\,.
\ee
$R(b)$ is, therefore, logarithmically [multiplied by
$(1-3b)^2$] singular at $b=1/3$, but bounded.

In terms of $R(b)$ the constant is given by
\be
C=- \frac{\pi}{9} R\left(\frac{1}{3}\right)
\label{residue} \,.
\ee
The essential point for the perturbative calculation of the residue is that
the right-hand side of Eq. (\ref{residue}) 
can be written as a convergent series.
The perturbation series for $R(b)$,
\be
R(b)=\sum_{n=0}^\infty r_n b^n\,
\label{residue-series}
\ee
is convergent on the disk $|b| \leq 1/3$ (note the boundary is included).
Being bounded, though singular, at $b=1/3$, $R(b)$ can be evaluated
in series at $b=1/3$.

We can now do some numerical checks to see how rapidly the series
(\ref{residue-series}),when evaluated at $b=1/3$,
converges to the known exact value.
From the perturbative coefficients  for $E(\alpha)$ given in
\cite{bpz} the coefficients $r_n$ in Eq. (\ref{residue-series}) can be
obtained. In Table \ref{table1}
we give the first terms of $C_{{\rm N}}$
defined by
\be
C_{{\rm N}} =-\frac{\pi}{9} \sum_{n=0}^{{\rm N}} r_n \left(
\frac{1}{3}\right)^{n}\,.
\ee
Note that $C_\infty=1$.
The numbers show that from $C_1$ to $C_5$ the series approaches the
true value in a steady pattern.
\begin{table}
\caption{\label{table1}
Sum of the first $N+1$ terms of the perturbation series
for the normalized residue ($C_\infty=
\tilde C_\infty=1$).}
\begin{ruledtabular}
\begin{tabular}{ccccccc} 
N  & 0 &1 &2 &3 &4 &5 \\  \hline
$C_{{\rm N}}$& 0.349 & 0.175 & 0.339 &0.487 
&0.631&0.759\\ \hline
$\tilde C_{{\rm N}}$& 0.349 & 0.109 & 0.502 
& 0.650 & 0.862 &0.994
\end{tabular}
\end{ruledtabular}
\end{table}

Since
the positions of the singularities for $\tilde E(b)$
are known we can improve the convergence using an ``optimal'' conformal
mapping \cite{mueller4}
\be
w=w(b) \,.
\ee
An optimal mapping for our case is that
$R(b(w))$ become as smooth as possible within the convergence disk
of the perturbation expansion in the $w-$plane
\be
R(b(w))=\sum_{n=0}^\infty \tilde r_n w^n \,.
\ee

An obvious strategy for an optimal mapping is to push  
all other singularities in the Borel plane
except for the first one  far away from
the origin. Here we consider a mapping
\be
w=\frac{1-\sqrt{1-3b/2}}{1+\sqrt{1-3b/2}}\,.
\ee
This maps the first singularity to $w=w_0$, where
\be
w_0= \frac{\sqrt{2}-1}{\sqrt{2}+1} \approx 0.171 \,,
\ee
and all other singularities to the unit circle.
Because the singularities (other than the first one) 
in the  $w-$plane are relatively farther away from the origin
than in the  $b-$plane, we expect
the series in  the $w-$plane to give better convergence.
In fact, the first terms shown in Table \ref{table1} 
of $\tilde C_{{\rm N}}$, which is defined by
\be
\tilde C_{{\rm N}} =-\frac{\pi}{9} \sum_{n=0}^{{\rm N}} 
\tilde r_n w_0^{n}\,,
\ee
show  a definite improvement in convergence (note again $\tilde C_\infty=1$). 

The result of this exercise shows that the leading term in the
nonperturbative 
effect caused by the instanton--anti-instanton  pairs on the ground 
state energy can be calculated
accurately (99\% accuracy) with only the first six terms of
the perturbation series for the ground state energy.

\section{The two-dimensional $O(N)$ nonlinear sigma model}

This model is exactly solvable in $1/N$ expansion and mimics
many interesting features of quantum chromodynamics (QCD).
It has an asymptotic freedom, dimensional transmutation, and
most interestingly for us the infrared (IR)  and
ultraviolet (UV) renormalons at the next leading order in $1/N$
expansion. Moreover, it was through the  studies of this model
\cite{david1,david2,david3,nsvz1,nsvz2}
that the link between operator product expansion (OPE) and IR
renormalons  suggested by Parisi \cite{parisi} has become more transparent.
This model can also provide a nontrivial test of our
proposed mechanism.

The two-dimensional $O(N)$ nonlinear sigma model is defined by
the (Euclidean) action
\be
S = \frac{1}{2} \int d^2x \sum_{a=1}^{N} \left[ \partial_\mu\sigma^a(x)
\partial_\mu\sigma^a(x) +\frac{\alpha(x)}{\sqrt{N}}\left(\sigma^a(x)\sigma^a(x)
-\frac{N}{4\pi f}\right)\right]\,,
\ee
where $\alpha(x)$ is an auxiliary field, and $f$ is the coupling constant.
At the leading order in $1/N$ the $\sigma$ fields get dynamical mass
\be
m^2=\mu^2 e^{-1/f(\mu)} \,,
\label{mass}
\ee
where $\mu$ is the renormalization scale,
through the vacuum condensate of the auxiliary field
\be
\langle 0|\alpha(0)|0\rangle =-\sqrt{N}m^2 \,.
\ee
Equation (\ref{mass}) also defines the renormalization group (RG) running
of the coupling constant. The  $\beta$ function in the leading order in $1/N$
is therefore given by
\be
\beta(f)=\mu^2\frac{d f}{d \mu^2} =-f^2 \,.
\label{betafunction}
\ee

We shall now test our proposed mechanism with the truncated two-point function
of the $\sigma$ fields
\be
\Gamma(p^2)=p^2 +\Sigma(p^2) \,,
\ee
which is known to
all orders in OPE at order $1/N$ via the exact calculation of the
self energy $\Sigma(p^2)$ in \cite{bbk}.
$\Gamma(p^2)$ can be expanded in powers of $m^2/p^2$, which corresponds to
an OPE, as
\be
\Gamma(p^2)=p^2\left[ C_0(f(p)) +C_1 (f(p)) \frac{m^2}{p^2} +
O\left(\frac{m^4}{p^4}\right)\right]\,.
\label{ope0}
\ee
We keep here only the first two terms because
we are focused on the nonperturbative effect associated with the first IR
renormalon.
The first term contains the usual perturbation expansion,
and the second term, which comes from the vacuum condensate of
$\alpha(x)$, is the nonperturbative amplitude that gives rise
to the first IR renormalon. The terms of higher powers in $m^2/p^2$
are associated with the higher renormalon singularities, and shall be
ignored.

At the leading order in $1/N$, $C_0=C_1=1$. At order $1/N$ they have a rich 
structure and  read  \cite{bbk} 
\bear
C_0(f(p)\pm{\rm i}\epsilon)&=&\frac{1}{N} \int^{\infty \pm {\rm i} 
\epsilon }_{0 \pm{\rm i}\epsilon} db
\left[ e^{-b/f(p)} \left(\frac{1}{f(p)}
F^{(0)}(b) 
+G^{(0)}(b)\right) - H^{(0)}(b)\right] \nonumber \\
C_1(f(p)\pm{\rm i}\epsilon)&=&-\frac{1}{N} \int^{\infty \pm {\rm i} 
\epsilon }_{0\pm {\rm i} \epsilon } db
\left[ e^{-b/f(p)} \left(\frac{1}{f(p)} 
F^{(1)}(b)
+G^{(1)}(b)\right) - H^{(1)}(b)\right] \,,\nonumber\\ 
\label{c0c1}
\eear
where
\bear
F^{(0)}(b)&=&1 \nonumber \\
G^{(0)}(b)&=&\frac{1}{b} +\frac{1}{b-1} -\psi(1+b)-\psi(2-b) -2\gamma_{E}
\nonumber\\
H^{(0)}(b)&=& \frac{1}{b} +B_1(b) \,,
\eear
and
\bear
F^{(1)}(b)&=&b^2-1 \nonumber \\
G^{(1)}(b)&=&-\frac{1}{b} +2 b^2 -4 b+1+(1-b^2)\left[\psi(1+b)+\psi(2-b) 
+2\gamma_{E}\right]
\nonumber\\
H^{(1)}(b)&=&- \frac{1}{b} -\frac{1}{b-1} -\frac{1}{1+b} +B_0(b) \,.
\eear
$B_0(b), B_1(b)$, whose exact forms  are not
important for us, are analytic functions.

Note the  renormalon poles at $b=n$, where $n$ is a nonzero integer,
in $G^{(0)}, H^{(1)}$.
To avoid these poles
the integration contour can be either on the upper or the lower
half plane; for consistency, however, an identical contour should be
taken for $C_0$ and
$C_1$.

We shall identify the first term  in Eq. (\ref{ope0})
as the perturbative amplitude $\Gamma_{\rm PT}$
and the second term as the nonperturbative amplitude $\Gamma_{\rm NP}$.
Performing the integration over $b$  in Eq. (\ref{c0c1})
it is then easy to see that $\Gamma_{\rm NP}$ at order $1/N$ is
given by
\bear
N\Gamma_{\rm NP}(f(p)\pm {\rm i} \epsilon) &=&-
\frac{m^2}{p^2} \left[ \ln f(p) \mp {\rm i}\pi + {\rm 
real~ const.} + O(f)\right]
\label{imaginary-sigmamodel}\\
&=& -e^{-1/f(p)} \left[ \ln(-f(p)\mp{\rm i}\epsilon)
+{\rm real~ const.} +O(f)\right]
\label{np-sigmamodel}\,,
\eear
where in the last step we have used Eq. (\ref{mass}).
Thus to the leading order in weak coupling
\be
N\Gamma_{\rm NP}(f)=-e^{-1/f}\ln (-f)\,,
\ee
which
has a branch cut along the positive real axis in the coupling
plane, again in agreement with our proposed mechanism.
The origin of the imaginary part in Eq. (\ref{imaginary-sigmamodel})
lies with  the ambiguity in obtaining
the renormalized condensate, in this case the condensate of the auxiliary
field $\alpha(x)$,
from the dimensionally regularized condensate in $2+\epsilon$ dimension
\cite{david1,david2,david3}.
When the perturbative and the
nonperturbative amplitudes are added together 
this imaginary part is canceled by the imaginary part
in $\Gamma_{\rm PT}(f(p)\pm{\rm i}\epsilon)$ coming
from the pole at $b=1$ in $G^{(0)}(b)$.  

From Eq. (\ref{c0c1}) the perturbation expansion for
$\Gamma_{\rm PT}(f(p))$ can be
easily obtained
\be
\Gamma_{\rm PT}(f(p)) =\ln f(p) + {\rm const.} +\sum_{n=0}^{\infty}
a_n f(p)^{n+1}\,,
\label{seriesforApt}
\ee
with
\bear
a_0&=&-2\,,\nonumber\\
a_n&=&n! \left[ \left(1+(-1)^n\right) \zeta(n+1) -2\right] \quad
\left({\rm for~} n\geq 1\right) \,,
\label{coefficients}
\eear
where $\zeta$ denotes the Riemann $\zeta-$function.
The logarithmic term in Eq. (\ref{seriesforApt})
arises from the anomalous dimension of
the $\sigma$ fields.

It is instructive to see now that the logarithmic dependence in
(\ref{imaginary-sigmamodel}) can be obtained
without knowing the exact solution for the $\sigma$ self-energy.
To see this let us make the renormalization scale dependence 
of $\Gamma$ explicit, which was hitherto 
implicitly suppressed.\footnote{This was allowable because 
the renormalization scale dependence
in $\Gamma(p^2)$ can be factored out.} 
$\Gamma(p^2,\mu^2,f(\mu))$ can be expanded in OPE as
\bear
\Gamma(p^2,\mu^2,f(\mu))&=&p^2[ \bar C_0(\mu^2/p^2,f(\mu)) +
\frac{1}{p^2}
\bar C_1(\mu^2/p^2,f(\mu))\langle 0|\alpha|0\rangle_{|_\mu}
\nonumber\\
&&+ {\rm higher~ dimension~
terms}]\,,
\label{ope}
\eear
where $\bar C_{\rm i}$ are the Wilson coefficients, and
the ignored terms involve operators of dimension four or higher.
From the RG equations for the coefficient $C_1(\mu^2/p^2,f(\mu))$ and
the condensate $\langle0|\alpha|0\rangle$ the second term in (\ref{ope})
can be easily written up to a momentum independent factor as
\be
\bar C_1(1,f(p)) e^{-1/f(p)}
\exp\left[-\int^{f(p)}[2\gamma_\sigma(f')+\gamma_\alpha(f')]/\beta(f')
df'\right]\,,
\ee
where $\gamma_\sigma(f)$ and $ \gamma_\alpha(f)$ denote 
the anomalous dimensions for the
$\sigma$ and $\alpha$ fields, respectively.
Comparing this term with the second one in Eq. (\ref{ope0})
we can identify
\be
C_1(f(p))=\bar C_1(1,f(p))  \exp\left[-\int^{f(p)} [2 \gamma_\sigma(f')
+\gamma_\alpha(f')]/\beta(f') df'\right]\,. 
\ee
Therefore the logarithmic term 
in Eq. (\ref{imaginary-sigmamodel}) can arise from a logarithmic term
at order $1/N$ 
in the Wilson coefficient $\bar C_1(1,f(p))$, and 
the anomalous dimensions 
$\gamma_{\{\sigma,\alpha\}}$, which are nonvanishing at order $1/N$:
\be
\gamma_{\{\sigma,\alpha\}}(f) \propto \frac{1}{N} [f +O(f^2)]\,.
\ee
What is noteworthy here is that the contributions from these sources
when added together
conspire to absorb the imaginary parts in Eq. (\ref{imaginary-sigmamodel})
into a cut-function.
If, for example, the logarithmic term had a different coefficient than
that given in Eq. (\ref{imaginary-sigmamodel}), then after the absorption of the
imaginary parts into a cut-function [$\ln(-f)$] there would have remained
a logarithmic piece ($\ln(f)$) which has a cut along the negative
real axis in the coupling plane. This would have failed our argument that
the nonperturbative amplitude has a cut only along the
positive real axis.

Now, once we know the form of the
anomalous dimensions for the $\sigma$ and $\alpha$ fields,
and the logarithmic dependence in
the Wilson coefficient at order $1/N$, 
the nonperturbative amplitude can be written with no help from
the exact solution for the $\sigma$ self energy as
\be
N\Gamma_{\rm NP}(f(p)) \propto e^{-1/f(p)}[ \ln f(p) + 
{\rm subleading~ terms}] \,.
\label{np-inferred}
\ee
Then demanding $\Gamma_{\rm NP}(f)$
have a branch cut along the positive real
axis in the coupling plane we can refine Eq. (\ref{np-inferred})
to
\be
N\Gamma_{\rm NP}(f(p)) = C e^{-1/f(p)}[ \ln(-f(p)) + 
{\rm subleading~ terms}]
\label{np-refined}\,,
\ee
with $C$ an undetermined real constant,
from which we can obtain  a relation valid in leading order in weak coupling:
\bear
{\rm Re}[\Gamma_{\rm NP}(f)]&=&\mp
\frac{1}{\pi} {\rm Im}[
\Gamma_{\rm NP}(f\pm{\rm i}\epsilon)]\ln
f(p)\nonumber \\
&=&\pm\frac{1}{\pi} {\rm Im}[ \Gamma_{\rm PT}(f\pm {\rm i}\epsilon)]
\ln f(p)\,.
\eear
Because the imaginary part of the
perturbative amplitude can be calculated from Borel resummation this relation
renders the real
part of the nonperturbative amplitude to be  calculable from the
perturbation  theory.

The leading term in $\Gamma_{\rm PT}(f)$ depends only on the
constant $C$. We can calculate this constant from perturbation theory
using the method  used in the previous section for a similar purpose.
First, we note that for the $\Gamma_{\rm PT}(f)$
to cancel the imaginary part coming from Eq.
(\ref{np-refined}) at positive $f(p)$
the Borel transform of $\Gamma_{\rm PT}(f)$  should have the
following singularity at $b=1$ (the first IR renormalon):
\be
\tilde \Gamma_{\rm PT} (b) = \frac{C}{1-b} \left[1 +O(1-b)\right] \,.
\ee
This shows that the constant $C$ becomes  the residue of
the renormalon singularity and can be written as \cite{tlee1,tlee2}
\be
C= R(1)
\ee
with
\be
R(b)\equiv (1-b) \tilde \Gamma_{\rm PT}(b)\,.
\ee

Because of the UV renormalon at $b\!=\!-1$  the residue
$C$ cannot be directly evaluated by
the perturbation expansion of $R(b)$ around the origin.
To map away other renormalons  than the first IR renormalon
we introduce a conformal mapping
\be
w=\frac{\sqrt{1+b}-\sqrt{1-b/2}}{\sqrt{1+b}+\sqrt{1-b/2}}
\ee
which sends the first IR renormalon to $w=1/3$ and
all other renormalons  to the unit circle.
Because in the $w-$plane the first IR renormalon is the
closest singularity to the origin,
$C$ can now be evaluated by plugging $w=1/3$ into the following
series expansion
of $R(b(w))$:\footnote{More precisely,
it is the expansion of $R(b)$  of the Borel transform for
 [$\Gamma_{\rm PT}(f)-\ln f -{\rm const.}$]
in Eq. (\ref{seriesforApt}) rather than that of the Borel transform for
$\Gamma_{\rm PT}(f)$.
Either of the Borel transforms can be used because both have an
identical renormalon singularity at $b=1$.}
\be
R(b(w))=\sum_{n=0}^{\infty} r_n w^n \,.
\ee

The first terms of $C_N$, which are defined by
\be
C_{N}=\sum_{n=0}^{N} r_n \left(\frac{1}{3}\right)^n \,,
\ee
were calculated using the perturbative  coefficients
(\ref{coefficients}).
The numbers in Table \ref{table2} show that the residue can be 
 determined with  good accuracy from the first terms of
the perturbation theory.
\begin{table}
\begin{ruledtabular}
\caption{\label{table2}
Sum of the first $N+1$ terms of the perturbation series
for the renormalon  residue ($C_\infty=-1$).}
\begin{tabular}{ccccccc} 
N  & 0 &1 &2 &3 &4 &5 \\  \hline
$-C_{{\rm N}}$& 2.000 & 2.000 & 0.100 &0.945 
&0.916&1.052
\end{tabular}
\end{ruledtabular}
\end{table}
 
\section{The QCD condensate effects}

We now come to the potentially most interesting application
of our proposed mechanism.
Because of the nonperturbative nature of the QCD vacuum,
operator condensates appear ubiquitously in low energy QCD
phenomenology, especially in the Shifman-Vainshtein-Zakharov (SVZ)
sum rule formalism
\cite{svz-sumrule1,svz-sumrule2}. The effects
of these condensates become stronger at lower energies
and become phenomenologically more important. They are
in general not calculable, and  treated as free parameters to be
fitted by experimental data.
The condensates are generally introduced through OPE. In Borel resummation
of the QCD perturbation theory they appear as the nonperturbative amplitudes
which are required to cancel the imaginary parts 
arising from the IR renormalon singularities.
The purpose of this section is to see the implication of our proposed 
mechanism  on the nonperturbative effects caused by these condensates.

In general the form of a nonperturbative effect  due to 
the condensates can be determined by
OPE and the RG equations for the associated  Wilson coefficients and the
condensates.
Once the form is determined then we can further refine it
by demanding the nonperturbative amplitude have a branch cut only
along the positive real axis in the coupling plane.
This will then allow us to write the real part of the amplitude 
in terms of the imaginary part that can be calculated from
Borel resummation.

Of course, one should remember that not all nonperturbative effects
in QCD could be related to perturbation theory. It is clear, for example,
that perturbation theory cannot have any bearing on the nonperturbative 
effects arising in chirality violating processes. 

For definiteness, we shall consider the Adler $D$-function in massless QCD
defined
by 
\be
D(Q^2)=-4\pi^2 Q^2 \frac{d\Pi(-Q^2)}{ dQ^2}-1\,,
\ee
where $\Pi(q^2)$ ($q^2\equiv -Q^2$)
is the vacuum polarization function in the Euclidean region ($q^2 <0$)
of the current $j^\mu(x)=
\bar{u}(x)\gamma^\mu d(x)$, with $u ,d$ denoting the {\it up} and {\it down}
quarks.
$D(Q^2)$ can be expanded in OPE as 
\bear
D(Q^2)&=& C_0(Q^2) +C_4(Q^2)\frac{\langle0|O_4|0\rangle}{
Q^4} \nonumber \\
&&+{\rm higher~ dimension~ terms}\,.
\label{ope-adler}
\eear
As before we focus on the nonperturbative effect associated with the
closest singularity to the origin on the positive real axis
in the Borel plane, in this case the first IR renormalon,
and ignore terms of dimension six or higher
since they are associated with the higher IR renormalons.
$\langle0|O_4|0\rangle$ is the renormalization scale invariant
gluon condensate of the  dimension-four operator
\be
O_4\equiv -\frac{1}{\pi\beta_0}\left[\frac{\beta(\alpha_s)}{\alpha_s}
G_{\mu\nu}^a G^{a\mu\nu}\right]\,,
\ee
where $G_{\mu\nu}^a$ is the gluon field strength tensor, 
and $\alpha_s$ is the strong coupling constant.
$\beta(\alpha_s)$ is the QCD $\beta-$function:
\be
\beta(\alpha_s)=\mu^2\frac{d \alpha_s(\mu)}{d\mu^2}=
-\alpha_s^2\left[\beta_0 +\beta_1 \alpha_s
+O(\alpha_s^2)\right]\,,
\label{qcd-betafunction}
\ee
where for $N_c$ colors and  $N_f$ quark flavors
\bear
\beta_0&=&\frac{1}{4\pi}\left(
\frac{11}{3}N_c-\frac{2}{3}N_f\right)\,,\nonumber\\
\beta_1&=&\frac{1}{(4\pi)^2}\left(
\frac{34}{3}N_c^2-\frac{N_c^2-1}{N_c}N_f -\frac{10}{3}N_c N_f\right)\,.
\eear
The Wilson coefficients $C_0, C_4$  can be expanded in
power series in the strong coupling $\alpha_s(Q)$:
\bear
C_0(Q^2)&=&\sum_{n=0}^\infty d_n^{\left(0\right)} \alpha_s(Q)^{n+1}
\label{series-adler}\,,\\
C_4(Q^2)&=&\frac{2\pi^2}{3}\left(1+
\sum_{n=1}^\infty w_n^{\left(0\right)} \alpha_s(Q)^n \right)\,,
\label{wilson-adler}
\eear
where the coefficients $d_n^{\left(0\right)}, w_n^{\left(0\right)}$
are real numbers.

We shall now identify the first term in Eq. (\ref{ope-adler})
as the perturbative amplitude $D_{\rm PT}(\alpha_s(Q))$
and the second term of gluon condensate as the
nonperturbative amplitude $D_{\rm NP}(\alpha_s(Q))$.
$D_{\rm PT}$ can be expressed as a Borel resummation
of the perturbation series (\ref{series-adler})
\be
D_{\rm PT}(\alpha_s(Q) \pm {\rm i}\epsilon)=
\frac{1}{\beta_0}
\int^{\infty\pm {\rm i}\epsilon}
_{0\pm{\rm i}\epsilon} db\, e^{-b/\beta_0\alpha_s(Q)} 
\tilde D_{\rm PT}(b)\,,
\label{resummation-qcd}
\ee
where $\tilde D_{\rm PT}(b)$ is defined by
\be
\tilde D_{\rm PT}(b)=\sum_{n=0}^{\infty}
\frac{d_n^{\left(0\right)}}{n!} \left(\frac{b}{\beta_0}\right)^n \,.
\label{boreltransform-qcd}
\ee
This series is expected to have a finite radius of convergence ($|b|\!=\!1$)
set by the UV renormalon at $b=-1$. Beyond the radius of convergence
$\tilde D(b)$ is assumed to be obtained by analytic continuation.
$D_{\rm PT}$ is now expected to have an imaginary part with
sign ambiguity 
due to the first IR renormalon at $b=2$ in the Borel integral.
This imaginary part is to be
canceled by the imaginary part from $D_{\rm NP}$.
Now $D_{\rm NP}$, the second term in Eq. (\ref{ope-adler}),
can be written as
\be
D_{\rm NP}(\alpha_s(Q))\propto
\alpha_s(Q)^{-\nu} e^{-2/\beta_0\alpha_s(Q)} \tilde C_4
(\alpha_s(Q),\beta_i,w_i^{\left(0\right)})
\label{np-adler}
\ee
via the weak coupling expansion of
\be
\frac{\langle0|O_4|0\rangle}{Q^4} \propto
\exp{\left[-2\int^{\alpha_s(Q)}\frac{d\alpha'}{\beta(\alpha')}\right]}\,,
\label{condensate-ratio-qcd}
\ee
which comes from the RG invariance of  the gluon condensate.
Here,
 \be
\nu=2\beta_1/\beta_0^2\,,
\ee
which is  noninteger for most of the combinations of $N_c, N_f$.
$\tilde C_4$, which  comes from $C_4$ and the weak coupling expansion
of Eq. (\ref{condensate-ratio-qcd}), is real and calculable in perturbation,
and can be expanded
in power series:\footnote{When the 
higher renormalons are taken into account the series expansion
for $\tilde C_4$ is an asymptotic expansion, and should be Borel
resummed. However, since we ignore all higher renormalons
$\tilde C_4$ is assumed to be well defined by the series expansion.
The error on $\tilde C_4$ by this assumption is $O(e^{-1/\beta_0\alpha(Q)})$
which is due to the dimension six condensates.}
\be
\tilde C_4(\alpha_s(Q),\beta_i,w_i^{\left(0\right)})=1+
\tilde w_1^{\left(0\right)} \alpha_s(Q) +\tilde w_2^{\left(0\right)}
\alpha_s(Q)^2+\cdots\,,
\label{wilson-coefficient-modified}
\ee
where $\tilde w_n^{\left(0\right)}$ depends on $\beta_{i+1}$ and
$w_i^{\left(0\right)}$, $i \leq n$, respectively,
in  Eqs. (\ref{qcd-betafunction}) and (\ref{wilson-adler}).

Following the argument in Sec. \ref{sec2} we demand the imaginary part in
$D_{\rm NP}$ come from a branch cut along the
positive real axis in the  coupling plane.
Presence of UV renormalons, which gives rise to a sign alternating
large order behavior, does not affect this requirement , since
UV renormalons can be mapped away using a conformal mapping 
from the Borel integration contour,
thus causing no essential problem
for Borel resummation \cite{parisi,mueller}.
Since in the nonperturbative amplitude (\ref{np-adler})
a branch cut can arise only from the factor $\alpha_s(Q)^{-\nu}$,
with $\nu$ a noninteger number,
we conjecture that the nonperturbative amplitude is given
in the form:
\be
D_{\rm NP}(\alpha_s(Q))=
C \,[-\alpha_s(Q)]^{-\nu} e^{-2/\beta_0\alpha_s(Q)}
\tilde C_4 (\alpha_s(Q),\beta_i,w_i^{\left(0\right)})\,,
\label{np-adler-refined}\ee
where $C$ is an undetermined, dimensionless, real constant.
 In this specific case the power of the coupling
constant in the preexponential factor 
depends on $\nu$ only. But, in general the power
depends not only on $\nu$ but also on the one-loop anomalous dimensions
of the associated operators. In such a case, we propose that the branch
cut arises likewise from the preexponential power term in the coupling, and
the correct form for the nonperturbative amplitude can be obtained
by flipping the sign of the coupling constant
in the preexponential factor.

The argument leading to Eq. (\ref{np-adler-refined}) shows that when $\nu$
takes an integer value for a particular combination of $N_c$ and $ N_f$,
the nonperturbative amplitude $D_{\rm NP}$
cannot have an imaginary part. This implies disappearance of the
first IR renormalon singularity in the Borel plane for an integer $\nu$.
A further comment on this point will follow shortly.

Now with Eq.~ (\ref{np-adler-refined}) we have the real and the imaginary
parts,
\bear
{\rm Re}[D_{\rm NP}(\alpha(Q) \pm {\rm i} \epsilon)]
&=& C\,\cos(\nu\pi) \alpha_s(Q)^{-\nu} e^{-2/\beta_0\alpha_s(Q)}\tilde C_4,
\label{realpart-qcd}\\
{\rm Im}[D_{\rm NP}(\alpha(Q)\pm {\rm i}\epsilon)]
&=& \pm C\,\sin(\nu\pi)\alpha_s(Q)^{-\nu}e^{-2/\beta_0\alpha_s(Q)}
\tilde C_4\,,
\label{imaginarypart-qcd}
\eear
from which a relation between the real and the imaginary parts is 
obtained:
\be
{\rm Re}[D_{\rm NP}(\alpha_s(Q)\pm {\rm i}\epsilon)]=
\pm \cot(\nu\pi)
{\rm Im}[D_{\rm NP}(\alpha_s(Q)\pm {\rm i}\epsilon)] \,.
\label{qcd-relation}
\ee
This has an important implication.
It relates the usually incalculable real part of the nonperturbative
amplitude to its imaginary part
that is calculable from Borel resummation. Moreover, this relation
is not for some part only of the amplitude as in the previous
two examples, but holds to all orders in perturbative expansion
of $\tilde C_4$. This implies that as far as the gluon condensate
effect is concerned the nonperturbative effect
can be calculated completely from the Borel resummation
of the  perturbation theory.

From Eq.~(\ref{qcd-relation})
and the fact that the imaginary parts in $D_{\rm
PT}(\alpha_s(Q)\pm{\rm i}\epsilon)$ and $D_{\rm
NP}(\alpha_s(Q)\pm{\rm i}\epsilon)$ cancel each other, we can
write the Adler function in terms of $D_{\rm PT}$ only:
\be
D(Q^2)=\left[{\rm Re}
\mp \cot(\nu\pi) {\rm Im}\right] D_{\rm PT}(\alpha_s(Q)\pm {\rm
i}\epsilon)\,.
\ee
Thus both the real and the imaginary parts of the Borel resummation
are required to rebuild the true amplitude from the
perturbation theory.
The imaginary part comes from the region beyond the first IR renormalon
($b\geq 2$), and for its calculation analytic continuation
to the region beyond the radius of convergence
of the perturbative Borel transform (\ref{boreltransform-qcd})
is required. A more convenient method, though equivalent to the analytic
continuation, is to use a conformal mapping
to map, for example, all the renormalon singularities, or all the renormalon
singularities except for the
first IR renormalon, to the
unit circle. A conformal mapping of the first kind was used 
in rebuilding the imaginary part of a metastable-vacuum energy 
in a quantum mechanical model \cite{sy}, and the second kind was recently used
in Borel resummation of the real part of the
Adler function \cite{gl}, which may also be used for
the calculation of the imaginary part.
Here, instead, we shall try to evaluate the constant $C$
which would give a rough estimate of the nonperturbative
amplitude $D_{\rm NP}(\alpha_s(Q))$.

As in the previous examples, this constant becomes the residue,
up to a calculable constant, of the Borel transform at the
renormalon singularity at $b=2$.
In fact, for the Borel resummation (\ref{resummation-qcd})
to have an imaginary part that can cancel
the imaginary part (\ref{imaginarypart-qcd})
the Borel transform $\tilde D_{\rm PT}(b)$
should have the following singularity at $b=2$:
\bear
\tilde D_{\rm PT}(b)&=& \frac{C}{\Gamma(-\nu)}
(\beta_0/2)^{1+\nu}
(1-b/2)^{-1-\nu}\left[1 + O(1-b/2)\right] \nonumber\\
&&+{\rm
Analytic~ part}\,,
\label{renormalon-qcd}
\eear
where ``Analytic part'' denotes terms that are analytic at $b=2$.
In general the analytic part cannot be calculated.
Note that, as previously mentioned,
the renormalon singularity disappears
when $\nu$ takes an integer value. This is obvious for a negative integer
$\nu$, which happens, for example, at  $N_c=2, N_f=8$ with $\nu=-10$, or
at $N_c=3, N_f=15$ with $\nu=-176$. The disappearance of the renormalon
singularity for the latter case was noticed before \cite{tm}.
What seems to have been unexpected is 
that the singularity also disappears for a
non-negative integer
$\nu$, for example, at $N_c=6,N_f=12$ with $\nu=1$.
In this case the singularity disappears because
of the vanishing residue. The residue  vanishes 
since the constant $C$ should always be bounded, for the nonperturbative
amplitude (\ref{np-adler-refined}) cannot be divergent. 

With Eq.~(\ref{renormalon-qcd}), $C$  can now be obtained by
\cite{tlee1,tlee2}
\be
C=\Gamma(-\nu)(2/\beta_0)^{1+\nu} R(2)
\ee
with
\be
R(b)=(1-b/2)^{1+\nu} \tilde D_{\rm PT}(b)\,.
\label{rb-qcd}
\ee

As in the previous examples $R(2)$ may be evaluated as a perturbation series.
However, it is unlikely to obtain a good estimate of the constant $C$
by directly following the procedures in the previous examples,
since too few perturbative coefficients are known. Only the first
three terms of  the perturbation series for the Adler function are
presently known.
To improve the situation we will exploit the renormalization scale
independence of the Adler function. To do this, we replace
in Eqs.~(\ref{series-adler})--(\ref{resummation-qcd}), 
and (\ref{wilson-coefficient-modified})
the  coupling $\alpha(Q)$ with the running coupling
$\alpha(\xi Q)$, with $\xi$ defined by the renormalization scale
$\mu^2\!=\!\xi^2 Q^2$.
Then the perturbative coefficients in these equations should
change accordingly as $d_n^{\left(0\right)} \to d_n(\xi)$,
$w_n^{\left(0\right)} \to w_n(\xi)$, and $\tilde
w_n^{\left(0\right)} \to \tilde w_n(\xi)$, with
$d_n(\xi\!=\!1)\!=\!d_n^{\left(0\right)}$, 
$w_n(\xi\!=\!1)\!=\!w_n^{\left(0\right)}$,
and $\tilde w_n(\xi\!=\!1)\!=\!\tilde w_n^{\left(0\right)}$.
Also the Borel transform (\ref{boreltransform-qcd}) should be
redefined as
\be
\tilde D_{\rm PT}(b,\xi)=\sum_{n=0}^{\infty}
\frac{d_n(\xi)}{n!} \left(\frac{b}{\beta_0}\right)^n \,.
\label{boreltransform-qcd-modified}
\ee
Now the renormalization scale invariance of the gluon condensate
allows us to rewrite Eq. (\ref{np-adler-refined}) as
\be
D_{\rm NP}(\alpha_s(\xi Q))=
C\xi^4 \,[-\alpha_s(\xi Q)]^{-\nu} e^{-2/\beta_0\alpha_s(\xi Q)}
\tilde C_4 (\alpha_s(\xi Q),\beta_i,w_i(\xi))\,,
\label{np-adler-refined-modified}
\ee
and consequently the renormalon singularity corresponding to
Eq. (\ref{renormalon-qcd}) is given by
\bear
\tilde D_{\rm PT}(b,\xi)&=& \frac{C \xi^4}{\Gamma(-\nu)}
(\beta_0/2)^{1+\nu}
(1-b/2)^{-1-\nu}\left[1 + O(1-b/2)\right] \nonumber\\
&&+{\rm Analytic~ part}\,.
\label{renormalon-qcd-modified}
\eear
Note that $C$ in Eqs. (\ref{np-adler-refined-modified}) and
(\ref{renormalon-qcd-modified})
is the same one defined in Eq.
(\ref{np-adler-refined}), and is independent of the scale $\xi$
but dependent on the renormalization scheme.

With Eq. (\ref{renormalon-qcd-modified}), $C$ can be written as
\be
C=\frac{1}{\xi^4}\Gamma(-\nu)(2/\beta_0)^{1+\nu}R(b\!=\!2,\xi)
\ee
with
\be
R(b,\xi)=(1-b/2)^{1+\nu} \tilde D_{\rm PT}(b,\xi) \,.
\ee
We now proceed to evaluate $R(b\!=\!2,\xi)$ as a perturbation series.
Because of the UV renormalon at $b\!=\!-1$ the evaluation point $b\!=\!2$ is
beyond the convergence radius of the series for $R(b,\xi)$ around the origin,
hence we have to map away all other renormalons except for
the first IR renormalon, 
using a conformal mapping like
\be
w=\frac{\sqrt{1+b}-\sqrt{1-b/3}}{\sqrt{1+b}+\sqrt{1-b/3}}\,,
\ee
which maps the first IR renormalon to $w\!=\!1/2$ and all
other renormalons to the unit circle.
Now in the $w-$plane $C$ can be expressed as a convergent power
series, that is, $C\!=\!C_\infty$ where $C_N$  is defined by
\be
C_N(\xi^2)= \frac{1}{\xi^4}\Gamma(-\nu) (2/\beta_0)^{1+\nu} \sum_{n=0}^N
r_n(\xi)\left(\frac{1}{2}\right)^n
\ee
with $r_n(\xi)$ coming from the expansion
\be
R(b(w),\xi)=\sum_{n=0}^{\infty} r_n(\xi) w^n\,.
\label{r-series-qcd}
\ee
Although $C$ is $\xi$-independent, in general
$C_N$ will have a $\xi$-dependence because of its finite order summation.
This property is generic for any finite order QCD perturbation series and
can be used to improve the convergence of the series  by demanding
that at an optimal $\xi$ the scale dependence of the series be
minimal \cite{stevenson}.
Applying this idea to our problem we can hope that a better estimate
of the constant $C$ can be achieved by
taking $C_N(\xi^2)$ at an optimal $\xi_0$ at which
the unphysical $\xi$-dependence disappears locally,
\be
\left.\frac{d\, C_N(\xi^2)}{d\, \xi^2}\right|_{\xi=\xi_0}=0 \,.
\ee

Using the calculated next-next-leading order Adler function \cite{gkl,sa}
and the estimated $O(\alpha_s^4)$ coefficient \cite{gl,ks},
in the $\overline{\rm
MS}$ scheme at $N_f\!=\!3$ quark flavors ($\xi\!=\!1$)\footnote{
$d_n(\xi)$ in terms of $d_n^{\left(0\right)}$ can be found in \cite{gl}.}
\be
D(Q^2)=a(Q)  + 
1.6398\, a(Q)^2 +6.3710 \,a(Q) ^3 + d_3^{\left(0\right)} a(Q)^4 +O(a^5)\,,
\ee
where
$a(Q)\equiv \alpha_s(Q)/\pi$,
we give the last two calculable terms for $C_N$ in the $\overline{\rm
MS}$ scheme:
\be
C_2(1.7)=5.37\,,\quad C_3(2.2)=6.96 \,.
\label{cn-qcd}
\ee
\begin{figure}
 \includegraphics[angle=-90 
 ]{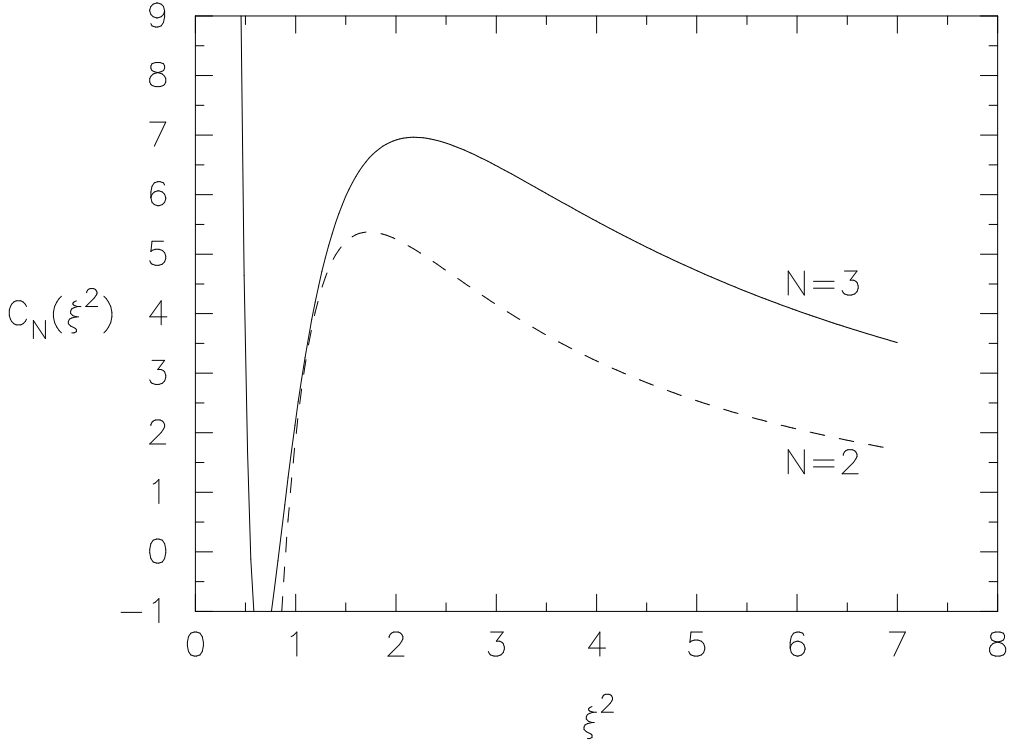}
\caption{\label{fig1} 
Renormalon residue vs renormalization scale $\xi^2$.}
\end{figure}
In this calculation we took  the estimated value $d_3^{\left(0\right)}
\!=\!25$,
which is from the
recent estimate using a technique called  
``bilocal expansion of Borel amplitude'' \cite{gl}.
This value is also in consistency with the well-known estimate in \cite{ks}.
Note that the optimal $\xi$ for $C_N$ is at $\xi_0^2\approx1.7$
for $C_2$ and $\xi_0^2\approx 2.2$ for $C_3$ (see Fig.~\ref{fig1}).

Because $C$ is evaluated with the perturbation
series (\ref{r-series-qcd}) at $w\!=\!1/2$ which is on the
boundary of the convergence disk,  $C_N$ in Eq. (\ref{cn-qcd})
should be regarded only as a rough estimate.
The speed of convergence of the sequence $C_N$ is expected to be
sensitive on the size  of the analytic part in Eq. 
(\ref{renormalon-qcd-modified}),
since the size of the singular term of $R(b,\xi)$ at $b\!=\!2$ is 
determined by this analytic part.

Using this estimate of $C$ we can now
evaluate the real part of the nonperturbative amplitude.
From Eqs.~ (\ref{realpart-qcd})
\bear
{\rm Re}[D_{\rm NP}(\alpha_s(Q)\!\pm\! {\rm
i}\epsilon)]&\approx&
C_3(2.2) \cos(\nu\pi) \alpha_s(Q)^{-\nu} e^{-2/\beta_0\alpha_s(Q)}\tilde C_4
\nonumber\\
&\approx& C_3(2.2) \cos(\nu\pi)\frac{ \Lambda_{{\overline {\rm MS}}}^4}{
Q^4}
\left[1\!+\! w_1^{\left(0\right)}\alpha_s(Q)\!+\! O(\alpha_s^2)\right]
\eear
where $w_1^{\left(0\right)}$ is defined in Eq. (\ref{wilson-adler}) and
we have substituted $C_3(2.2)$ for $C$,
 $\Lambda_{{\overline {\rm MS}}}$
denotes the ${\overline {\rm MS}}$ renormalization scale.

Now this nonperturbative amplitude can be translated to a
gluon condensate in the OPE (\ref{ope-adler}). We define the gluon condensate
by this nonperturbative amplitude by
\bear
\langle \frac{\alpha_s}{\pi}G_{\mu\nu}^a G^{a\mu\nu}\rangle_{\rm NP}
&\equiv&
\frac{3 Q^4}{2\pi^2}{\rm Re}[D_{\rm NP}(\alpha_s(Q)\pm {\rm
i}\epsilon)]\nonumber\\
&\approx& \frac{3}{2\pi^2}  C_3(2.2)\cos(\nu\pi)
\Lambda_{{\overline {\rm MS}}}^4
\nonumber\\
&\approx& 0.005\,\, {\rm GeV^4}
\label{condensate-estimate}
\eear
where
we have used $\Lambda_{{\overline {\rm MS}}}\approx 370 \,{\rm MeV}$
for $N_f\!=\!3$ quark flavors \cite{aleph}.

One should not compare the estimated value (\ref{condensate-estimate})
directly with the
phenomenologically fitted gluon condensate, for example, from the
QCD sum rule. In the QCD sum rule, the difference between
the Borel resummed perturbative amplitude and the sum of the first
terms in the perturbation series is approximated by power corrections,
and therefore the phenomenologically fitted condensate includes contributions
not only from the nonperturbative amplitude but also
from the perturbative amplitude. 
One may try to extract the gluon condensate effect in the
resummed perturbative amplitude through Eq. (\ref{resummation-qcd}),
for example,
by computing the minimal term of the perturbation series
with the large order behavior given by the renormalon singularity
(\ref{renormalon-qcd}). We believe, however, that this is not necessary, and
also not a good way to handle the renormalon effect.
A better approach to incorporate the renormalon effect, with only the first
few terms of the perturbation series available, is to write
using Eq. (\ref{rb-qcd}) 
the Borel transform $\tilde D_{\rm PT}(b)$ in the Borel integral
(\ref{resummation-qcd})
as
\be
\tilde D_{\rm PT}(b)=\frac{R(b)}{(1-b/2)^{1+\nu}}
\ee
and do perturbation on $R(b)$ instead of doing perturbation directly on
$\tilde D_{\rm PT}(b)$. This way the Borel transform 
can be better described in the most important region in the Borel
integral, i.e.,  between the origin and the first IR renormalon
singularity and the region just beyond the singularity \cite{gl}.

Nonetheless, it is interesting
to observe that the estimated value (\ref{condensate-estimate}) is
remarkably close to the recent estimate of gluon condensate
$0.006 \pm 0.012 \,{\rm GeV}^4$ \cite{ioffe} which was obtained by
fitting the spectral function of hadronic $\tau-$decay using  QCD sum rule.  

Finally, some comments are in order.
A full amplitude in QCD in general has infinitely many
cut singularities in the complex coupling plane as  
shown by 't~Hooft \cite{thooft}.
One may wonder how this can be compatible with our proposed mechanism
that is based on the proposition that the nonperturbative amplitude as
well as the Borel resummed amplitude
of perturbation theory have  cuts only along the positive real axes.
The resolution of this question lies probably with the nonconvergence
of the OPE. Since the OPE is expected to be an asymptotic
expansion \cite{shifman}, each term in the OPE, to the Wilson coefficient
of which the Borel resummation is applied, does not have to have the 
same singularities of the true amplitude.
Also, throughout this paper, the nonperturbative effects due to the 
higher dimensional operator
condensates were consistently ignored. We expect, however, there should
be no fundamental difficulty in incorporating them along the lines
described in this section. We conjecture that the nonperturbative effect
by the condensate of a dimension $2n$ operator in the Adler function
can be written as 
\be
D_{\rm NP}(\alpha_s) = C_n[-\alpha_s]^{-\nu_n} \exp(-n/\beta_0\alpha_s)
\tilde C_{2n} (\alpha_s)\,,
\ee
where $\nu_n$ is a constant calculable  from the RG equations on the
Wilson coefficient and the condensate, and 
$\tilde C_{2n}$ is a modified Wilson
coefficient defined in a similar fashion as the 
$\tilde C_4$ in Eq. (\ref{wilson-coefficient-modified}).
Here $\tilde C_{2n}$ should be Borel resummed 
in the manner described in this section.
The unknown constant $C_n$ can then be determined by demanding
that the imaginary part of order
$\alpha_s^{-\nu_n} \exp(-n/\beta_0\alpha_s)$  be canceled by
the imaginary parts coming from 
the amplitudes associated with the operators
of lower dimensions.

\section{Summary}

Based on the general argument on Borel resummation of a same sign 
perturbation series
we have argued that the nonperturbative effect associated with the
divergence of the perturbation series should have a branch cut only along
the positive real axis in the coupling plane.
Demanding that the nonperturbative amplitude have such a branch cut
constrains the  form of a part of the nonperturbative amplitude,
the part from which the branch cut arises,
sufficiently that a relation can be established  between
the usually incalculable
real part  of the nonperturbative amplitude and  the
imaginary part that is calculable from the Borel resummation.
This way part of the real part of the nonperturbative amplitude, which
usually includes the leading term in weak coupling expansion, can be
calculated from Borel resummation of the perturbation theory.

As a nontrivial test, this mechanism was applied to the ground state energy of
the double well potential and the two-point function
in the two-dimensional $O(N)$ nonlinear sigma model at order $1/N$.
In agreement with our proposed mechanism the nonperturbative amplitudes
in these models have branch cuts only along the positive real axis in the
coupling plane.
With this mechanism the leading terms  of the 
nonperturbative amplitudes in these models could be
calculated with good accuracy from the first terms of the
corresponding perturbation series.

We then applied this mechanism to
the QCD condensate effects, particularly
the gluon condensate effect, and suggested that 
some of the condensate effects can be calculated from
perturbation theory, and gave an estimate of the
nonperturbative amplitude induced by the gluon condensate
using the known perturbative calculations of the Adler function.
We  observed that this mechanism could be applied to QCD, 
despite the fact that 
a true QCD amplitude has an infinite number of cut singularities in
the coupling plane, since the OPE to which the Borel resummation is
applied is not a convergent expansion.

\begin{acknowledgments}
The author is thankful to Gorazd Cvetic for a
careful reading of the manuscript and useful comments. Part of the work
was completed while the author was visiting Universidad Tecnica Federico
Santa Maria, Valparaiso, Chile. The warm hospitality by the members of
the physics department is gratefully acknowledged. This work was 
supported by BK21 Core Project.

\end{acknowledgments}

\end{document}